# Characterization of multilayer stack parameters from X-ray reflectivity data using the PPM program: measurements and comparison with TEM results

D. Spiga[a*], A. Mirone[b], G. Pareschi[a], R. Canestrari[a], V. Cotroneo[a], C. Ferrari[c],
C. Ferrero[b], L. Lazzarini[c], D. Vernani[a]

[a] INAF/ Brera Astronomical Observatory, Via E. Bianchi 46, 23807 Merate (LC), Italy
[b] European Synchrotron Radiation Facility (ESRF), 6 rue Jules Horowitz, Grenoble, France
[c] IMEM-CNR Parco Area delle Scienze 37/A 43010 Fontanini- Parma, Italy

## ABSTRACT

Future hard (10 -100 keV) X-ray telescopes (SIMBOL-X, Con-X, HEXIT-SAT, XEUS) will implement focusing optics with multilayer coatings: in view of the production of these optics we are exploring several deposition techniques for the reflective coatings. In order to evaluate the achievable optical performance X-Ray Reflectivity (XRR) measurements are performed, which are powerful tools for the in-depth characterization of multilayer properties (roughness, thickness and density distribution). An exact extraction of the stack parameters is however difficult because the XRR scans depend on them in a complex way. The PPM code, developed at ERSF in the past years, is able to derive the layer-by-layer properties of multilayer structures from semi-automatic XRR scan fittings by means of a global minimization procedure in the parameters space. In this work we will present the PPM modeling of some multilayer stacks (Pt/C and Ni/C) deposited by simple e-beam evaporation. Moreover, in order to verify the predictions of PPM, the obtained results are compared with TEM profiles taken on the same set of samples. As we will show, PPM results are in good agreement with the TEM findings. In addition, we show that the accurate fitting returns a physically correct evaluation of the variation of layers thickness through the stack, whereas the thickness trend derived from TEM profiles can be altered by the superposition of roughness profiles in the sample image.

## 1. INTRODUCTION

Multilayer coatings will be adopted in a number of future X-ray telescopes (HEXIT-SAT[1], Con-X[2], XEUS[3], SIMBOL-X[4]...) in order to enlarge their reflectivity (at grazing incidence angles of 0.2 ÷ 1 deg) to the hard X-ray band, from 10 keV up to 70 - 80 keV. This approach, already considered and studied in previous works[5,6], is particularly attractive since it would allow the extension of performances and results of technologies already adopted in soft X-rays telescopes to hard X-rays, like the mandrel replication by Nickel electroforming, already successfully used for the soft, single Au layer optics of SAX[7], Newton-XMM[8], JetX-SWIFT[9]. Typical multilayers that could be used to cover a continuous, hard X-ray band up to 70-80 keV can be a stack of 100-150 bilayers of Pt/C or W/Si[10], the design of which is to be optimized according to the specific cases. Moreover, the response of hard X-rays multilayers for astronomical applications can be enhanced in the soft X-ray band, by overcoating them with an additional Pt or Ir layer plus a final capping Carbon layer[11]. Multilayer coatings are equally used to concentrate/deviate X-rays along synchrotron beamlines, as well as in neutron optics (Ni/Ti), and in EUV optics (Mo/Si) for nanolithography.

Design, development and characterization of multilayer samples along with optics prototypes are research tasks at INAF/Brera Astronomical Observatory in the framework of international collaborations addressed to the manufacturing of X-ray telescopes of the next future. Multilayer coatings can be deposited using different methods (e-beam evaporation, ion-assisted deposition, ion-etched deposition, DC or RF magnetron sputtering, ion-beam sputtering). Each process has to be optimized for the specific application. In all cases high precision in thickness determination and low

---

* contact author: e-mail daniele.spiga@brera.inaf.it, phone +39-039-9991146, fax +39-039-9991160

interfacial microroughness are required. For the manufacturing of X-ray optics for astronomical applications large surface coverage and deposition process repeatability are also needed. Moreover, the large mirror shell number to be produced requires the process implementation in an industrial production context.

In the framework of a preliminary project[12] funded by the *Italian Space Agency* deposition methods of flat, Pt/C and Ni/C coated multilayer samples by e-beam evaporation have been thoroughly investigated, since this technique ensures large surface coverage and high deposition rates. The aim of the project is to prove the feasibility of the design and production of hard X-ray telescopes using the existing industrial facilities. The results till now are very encouraging [13], as the peak reflectivity achieved with this method turned out to be very high (> 93%). One of the shortcomings of this method is the non-steady evaporation rate, especially when evaporating light elements like Carbon: this can cause instability of the layer thickness and a consequent deviation of the actual thickness distribution from the intended one. This problem is being resolved along with the achievement of steady evaporation conditions, as well as the interfacial roughness growth, which can be lowered by using an ion-assistance/polishing device. DC/RF magnetron sputtering is a well-known and wide spread technique since the deposition rate is very steady and the deposition process yields a little intrinsical roughness, even if the large surface coverage requires dedicated deposition facilities[14].

In all these cases, the progress in multilayer coating needs a tool to systematically assess the improvements of test samples (uniformity, smoothness, correct layer thicknesses in the stack): *Transmission-Electron-Microscope* (TEM) images allow a direct, quantitative evaluation of layers thickness, but due to the long, complex sample preparation procedure and the consequent sample destruction, it can be used for a limited number of samples. As a standard test for all samples, we routinely perform X-ray reflectivity (XRR) measurements at two standard energies: 8.05 keV and 17.45 keV with a Bede-D1 diffractometer. This method is non-destructive, quick, and does not require any particular sample preparation. Moreover, it is very suitable to characterize in-depth a multilayer mirror because it is particularly sensitive not only to TEM-visible properties (thickness, interdiffusion, crystallization state), but also to other properties (layers density, high-frequency microroughness) that cannot be measured by TEM.

XRR scans interpretation is nevertheless difficult, as it depends on a large number of parameters, namely all the thickness, density and roughness values in the stack. The recursive[15] or matrix[16] formalism (including the microroughness via a Fresnel coefficients modification[17] or the Nevot-Croce formalism[18]) may be used as a suitable base for fitting experimental XRR curves, but due to the large number of parameters which should be manually adjusted, one can just recover the average properties of the layers adopting a simple stack model. The exact XRR curve simulation would instead require to reproduce *all* the thickness irregularities in a N-bilayer stack, involving a manual adjusting of more than 4N independent parameters, including thickness and roughness of each layer. In order to handle the large number of parameters characterizing a multilayer and to extract the values that return the best fit of experimental XRR curves, a number of computer programs were developed, like PPM (*Pythonic Program for Multilayers*), which was developed by one of us (A. Mirone, ESRF) to perform a fast multi-parametric optimization of the reflectivity scans at one or more energies at the same time. Some preliminary PPM results were presented in a previous SPIE volume[19].

PPM has been tested on reflectivity scans of some multilayer samples (with almost constant d-spacing), deposited at the Media-Lario Techn. coating facility. The fitting capabilities of PPM have also been tested with a graded multilayer sample used as reflective coating for a hard X-ray optic prototype presented in this conference[20]. In section 2 we will show some properties of the Ni/C and Pt/C multilayer samples and their reflectivity curves: a brief description of PPM will be provided in section 3. The reflectivity measurements were performed at INAF/Brera Astronomical Observatory (Italy). The PPM results with these multilayers and the comparison with TEM images (taken at IMEM-CNR) analysis will be shown in section 4: the fit results and the application to a graded multilayer are briefly discussed in section 5.

## 2. MULTILAYER SAMPLES

For our analysis we have concentrated on test samples of almost-periodic multilayers deposited over flat Si wafers. The XRR scans of the chosen samples show a clear evidence of an irregular variation of d-spacing along the stack, and are therefore suitable to test the fitting capabilities of PPM. As the PPM fitting algorithm has to start from approximate, initial guess values of the average period, $\Gamma$ (heavier element thickness / d-spacing ratio) and $\sigma$ (roughness rms), we derived these values by a manual XRR simulation assuming first a stack structure described by a small number of parameters. The preliminary simulated XRR curves shown in fig. 1 and 2 were traced using the simulation routines of the IMD[21] package.

The first sample being considered is a Ni/C multilayer (19 bilayers) deposited at Media-Lario Techn. by e-beam evaporation, with the same facility used to deposit the single, Au reflective coating of SAX, XMM, SWIFT-XRT optics.

The reflectivity curves (at 8.05 and 17.45 keV), shown in fig.1, were already reported and analyzed by some of us in previous works[12,13,19]. After a laborious manual adjustment of parameters, the reflectivity angular scans calculated from a model have been brought to a discrete agreement with the experimental data, even though small reflectance features have not been fitted by the assumed model: the top reflectivity value of the first Bragg peak is very high (95% at 8.05 keV).

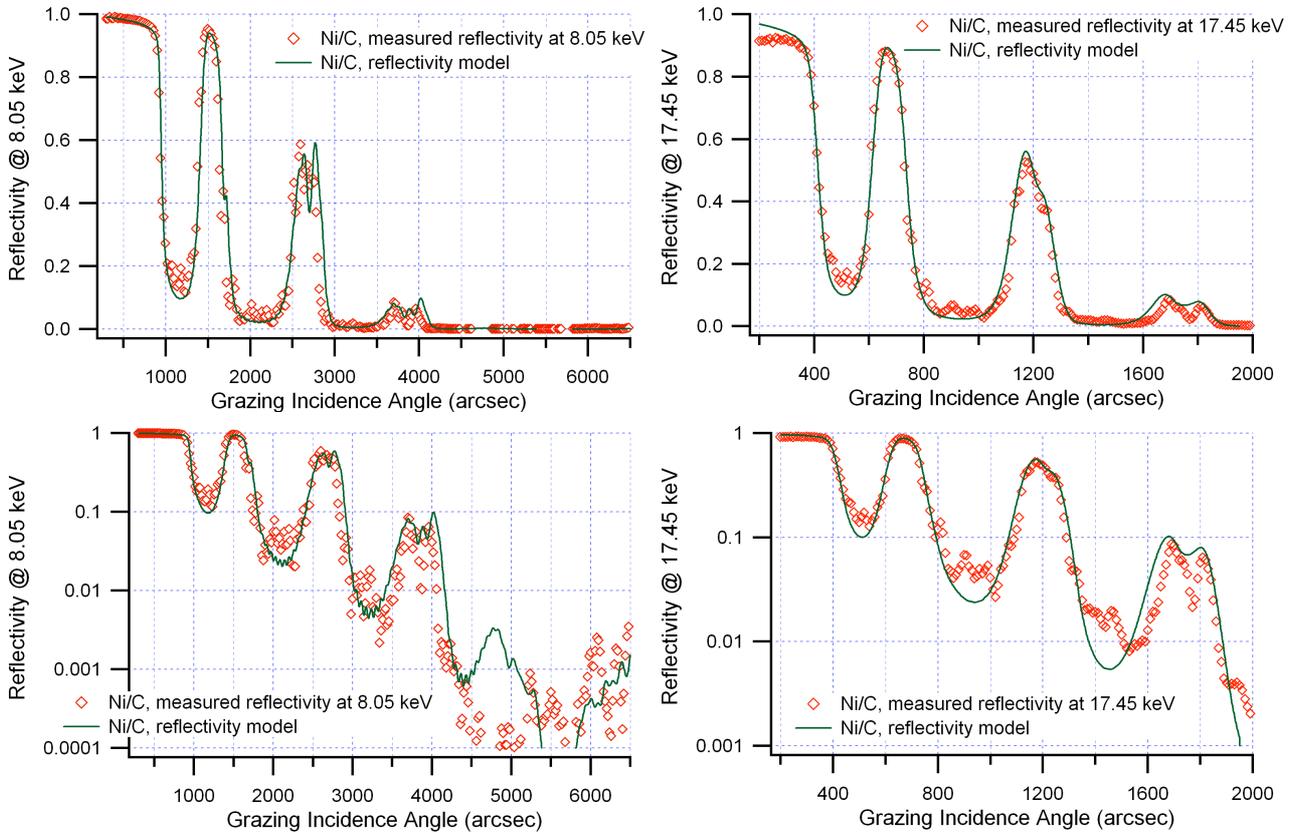

Fig.1: normalized XRR results (dots) at 8.05 (*left*) and 17.4 keV (*right*) of the Ni/C multilayer sample with 19 bilayers, in linear scale (*above*) and log scale (*below*): it is worthwhile noticing that the first Bragg peak reflectivity amounts to 95% @ 8.05 keV, a world record until 2005[22]; an uncontrolled period variation in the stack of about 2 nm is evident from the dispersion of the second peak. The solid line is the model, the parameters were derived manually.

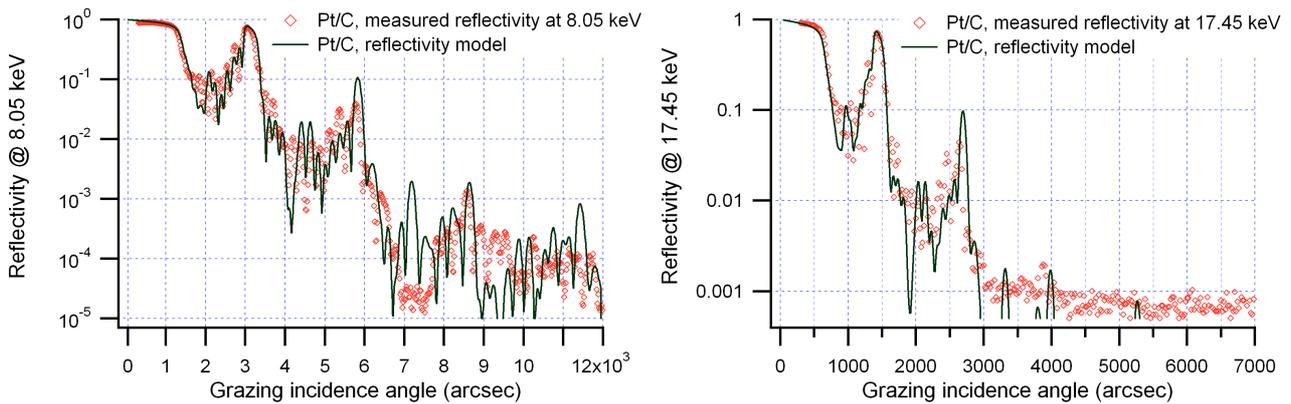

Fig.2: normalized XRR results (dots) at 8.05 keV (*left*) and at 17.45 keV (*right*) in log scale of the Pt/C multilayer test sample with 15 bilayers: the dispersion of d-spacing is quite evident from the complexity of the 2$^{nd}$ Bragg peak; from the manual adjustment we could not infer the drift of d-spacing. The linear plot is omitted because it does not show any relevant feature. The solid line is the model, the parameters were derived manually.

The manual fit-experiment matching suggests that there could be a d-spacing drift in the stack of about 2 nm across the 19 bilayers, to be probably ascribed to Carbon layers, even if this assumption does not explain many XRR features. The derived average period (13 nm) is much larger than the anticipated one (9 nm): the Ni layers thickness is approximately 3 nm. An interesting result of the fit is the Carbon density value (1.6 g/cm$^3$), much lower than the natural one (2.3 g/cm$^3$); Nickel has instead been found to be deposited at densities (8.7 g/cm$^3$) near the bulk value (8.9 g/cm$^3$). The roughness rms is good (4 Å with the actually found C density, 3 Å if one assumes the bulk density value) as we started from a Si wafer with a similar roughness value (3 Å).

The second sample is a Pt/C multilayer (15 bilayers) deposited by simple e-beam evaporation on a Si wafer: this sample is the result of a very important deposition test since it was used at INAF/OAB and Media-Lario Techn. to calibrate the deposition facility before producing a hard X-ray optic prototype[23]. The reflectivity scans (8.05 and 17.45 keV) are shown in figure 2. The same evaporation facility was used, but since the evaporation conditions had been calibrated and kept more stable, the average d-spacing turns out to be more correct (56 Å vs. 50 Å) with a $\Gamma$ factor around 0.42, as derived from the manual XRR modeling. Starting from the Silicon wafer roughness value ($\sigma_{rms} \approx 3$ Å), an average stack roughness of 4-5 Å rms fits well the data, as indicated by the good peak reflectivity (74% at 8.05 keV). The relevant dispersion of the 2$^{nd}$ and 3$^{rd}$ Bragg peak seems to be justified by a random variation rather than a continuous drift, but it cannot be estimated using a manual fit.

## 3. THE PPM COMPUTER PROGRAM

The analysis of the XRR curves presented in the previous section has been performed by using PPM, letting the fit parameters vary independently for each layer of the stack. The PPM code is specifically conceived for a fast evaluation of the stack parameters by fitting accurately reflectivity curves, even small reflectance features caused by the deeply buried layers: the thickness of layers can be measured within ~1 Å accuracy. The program reads in the reflectivity curves and a .xml source file defining the structural arrangement to be optimized. The number of necessary parameters may vary according to the supposed complexity of the stack. For example one can take each of the layer thicknesses as a free variable, or assume a gradual thickness drift. The other parameters are the layer densities and their roughness values. The user has also to set in the source file the lower and the upper limits and the initial guess value for every defined variable.

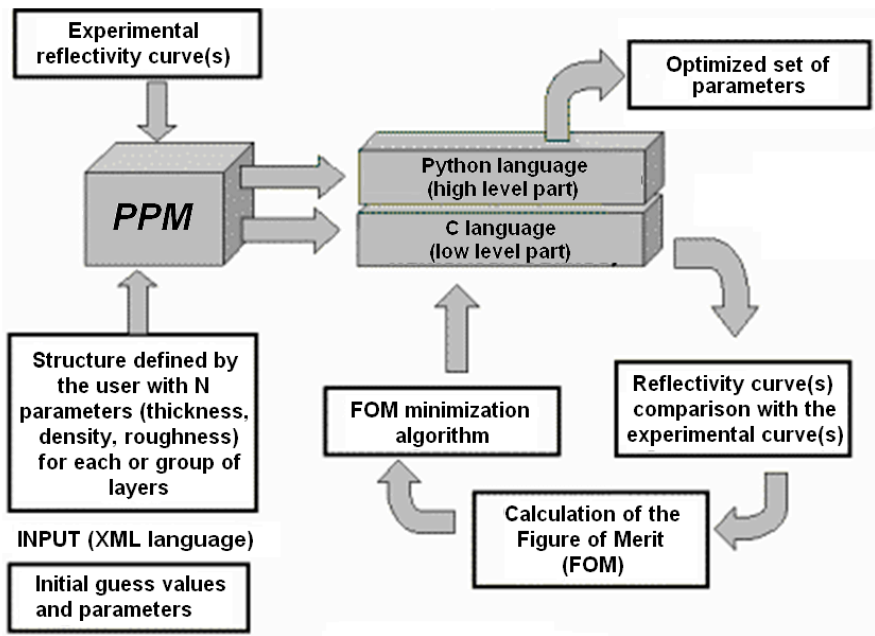

Fig.3: A scheme of the working principle of PPM: the reflectivity curves are initially calculated from the parameter values defined by the user for the chosen scheme which approximates the stack structure. The computed reflectivity (at one or more photon energies) is then compared with the experimental one(s). The Figure Of Merit (FOM) to be minimized is the $\chi^2$ of the logarithmic plots: the minimization algorithm (downhill - simulated annealing) varies the stack parameters within the limits fixed by the user. At every cycle the FOM is computed and the parameter set with the minimum FOM value is selected.

Starting from the initial values, PPM computes the reflectivity curve $R_c$ of the modeled multilayer structure at the photon energy of every available experimental scan, using the recursive theory of multilayers[15], with modified Fresnel coefficients to include the interfacial microroughness in the calculation[17]. The optical constants are computed from tabulated atomic scattering factors and from the actual material densities, that can also be assumed as fit parameters. Each computed curve is then compared with the measured reflectivity $R_m$, along with the Figure Of Merit (FOM):

$$FOM = \sum_i \frac{[\log R_m(i) - \log R_c(i)]^2}{|\log R_m(i)|}$$

where *i* is the index corresponding to the angular *i-th* sampled point.. Additional weights can be set in the FOM definition in order to improve the fitting in some particular sections of the XRR scan. The minimization algorithm varies the stack parameters within the limits fixed by the user. At every cycle the FOM is calculated and the parameter set with the minimum FOM value is selected . A flow chart of the PPM working scheme is shown in fig. 3.

It is well known that a logarithmic FOM assigns comparatively larger weights to small reflectivity values; hence, the automatic fit will be attracted also towards small reflectance features. The minimization algorithm implemented in PPM is a derivation of the well-known downhill simplex[3,24] (or amoeba) algorithm: this method represents the parameter set as a point in a *N*-dimensional space, in which a set of N+1 points moves following simple minimization rules, converging to the nearest FOM minimum: however, the downhill simplex alone would be very likely lead to a local minimum; hence very far from the global one. For this reason an annealing function[25,26,27], which is able to trigger a jump out of the local minimum, is implemented.

This simulated annealing function associates to every point of the parameter space a microscopic state of *N* particles and the convergence to a local minimum is compared to a thermalization process, leading to a state of definite temperature *T*. The system of particles is assumed to follow the Boltzmann distribution, where the FOM (depending on the actual particle state) plays the role of the energy. Using the downhill simplex alone, moves increasing the FOM would be always excluded. If the system stabilizes according to the Boltzmann distribution at a temperature large enough, there is instead a finite likelihood that the system does not converge to the local minimum.

Usually the initial temperature is set at a sufficiently high value for a large number of jumps out of the local minima to take place: while the system reaches the equilibrium, *T* is being slowly decreased, and the probability of increasing the FOM during a transition is lower and lower. At this point the system tends to converge towards the global minimum, till a minimum temperature value is reached, where the calculation stops. The probability of finding the global minimum is so increased, since a much larger portion of the parameter space could be explored with respect to the downhill simplex case, at the expense of an increase in the computation time.

The annealing procedure has also some critical points, such as the choice of the initial temperature and the cooling rate (usually an exponential decrease is adopted). A very slow cooling rate gives a greater degree of confidence that the global minimum has been found. These parameters, as well as the multilayer structure and the initial parameter values, have also to be carefully chosen in every single case. The reflectivity fitting can work with a large number of free parameters searching for the combination of values representing arguably the best possible fit of the scan. Once a set of values has been found, which returns a model reflectivity curve well superposing on the experimental one, one can be reasonably confident that this set represents a good approximation of the multilayer structure.

It should be noted, moreover, that a program using the simulated annealing procedure had already been developed for the optimization of graded multilayers at the ESRF[27].

## 4. PPM RESULTS OF XRR ANALYSIS

The fitting capabilities of PPM were assessed on the multilayers referred to in section 2. In order to reproduce short-term d-spacing irregularities, we have assumed in both cases each layer thickness to be a free variable, varying in a 3-4 nm wide interval: this very broad domain make the exploration of a very wide range of solutions possible. The calculations were first performed attributing data weights proportional to the reflectance values, in order to best fit to the primary Bragg peaks. The found values have then been used as initial guesses for the final calculation with no weights. Moreover, in order to refine the fit, the program needed to be restarted a few times, always using the values found in the previous run as initial points, until a stable solution was found.

**Pt/C multilayer**

The Pt/C (15 bilayers) multilayer reflectivity scans in fig. 2 were fitted simultaneously, considering the thicknesses of all layers as free parameters. Their values were left free to vary in the very wide intervals 15 - 40 Å for Pt and 20- 80 Å for C: initial values were 23 Å for Pt and 33 Å for Pt. Only the first C layer, deposited directly on the substrate, has been ignored since its presence does not affect noticeably the reflectivity profile. Roughness rms values for Pt and C were assumed to be constant in the stack, with variable values in the interval 2-6 Å. The densities have been fixed to values lower than the bulk densities (Pt: 20.5 g/cm$^3$ vs. 21.5 g/cm$^3$ - C: 1.6 g/cm$^3$ vs. 2.3 g/cm$^3$) as already experimented in depositing these elements deposited with the e-beam facility[12,13]. The final result matches very well the measured data both at 8.05 keV (see fig. 4) and 17.45 keV (see fig. 5): the fit improvement with respect to fig. 2 is apparent since the $\chi^2$ value of 44.9 at 8.05 keV has been reduced to 0.33. The $\chi^2$ at 17.45 keV has been also improved from 17.6 to 0.67.

The structure found by PPM to fit the measured reflectance values is shown in fig. 6 (left): the aperiodicity comes mostly from Carbon layers, with evident oscillations around 35 Å, and a peak at 75 Å corresponding to the 3$^{rd}$ C deposited layer. Platinum layers are more steady, oscillating around the value of 22 Å. It is worthwhile remembering that this fitting structure was found by assuming a regular, periodic multilayer as an initial guess. The derived roughness rms values are 5.3 Å for Pt and 4.2 Å for C.

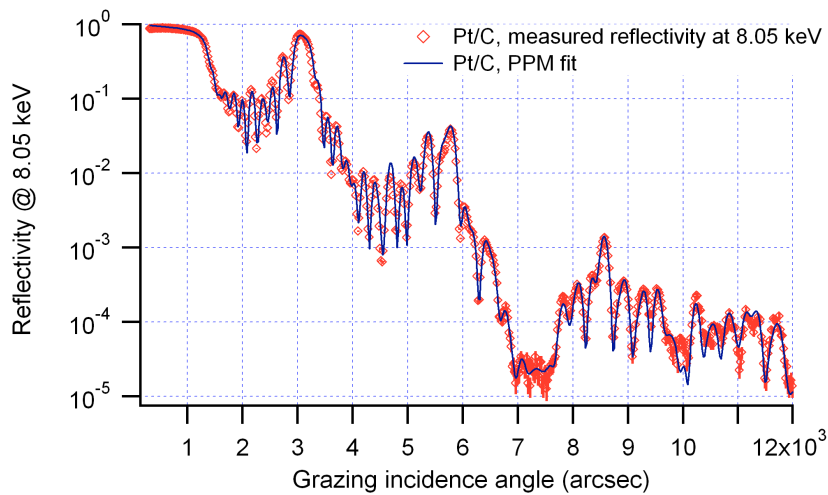

Fig. 4: the reflectivity at 8.05 keV of the Pt/C multilayer sample compared with the fitting model calculated via PPM (log scale). The fit procedure was run simultaneously on both scans at 8.05 and 17.45 keV (see fig. 5). Error bars (very small) are also included.

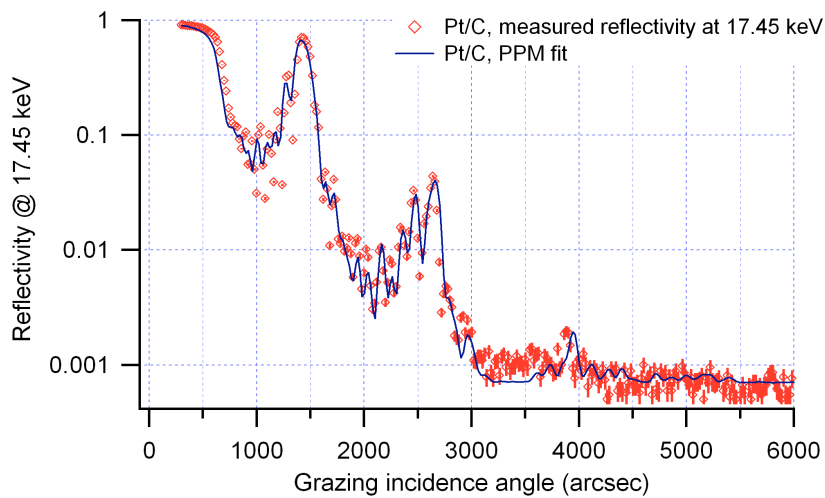

Fig. 5: the reflectivity at 17.45 keV of the Pt/C multilayer sample compared with the fitting model calculated via PPM (log scale). Error bars (very small) are also included.

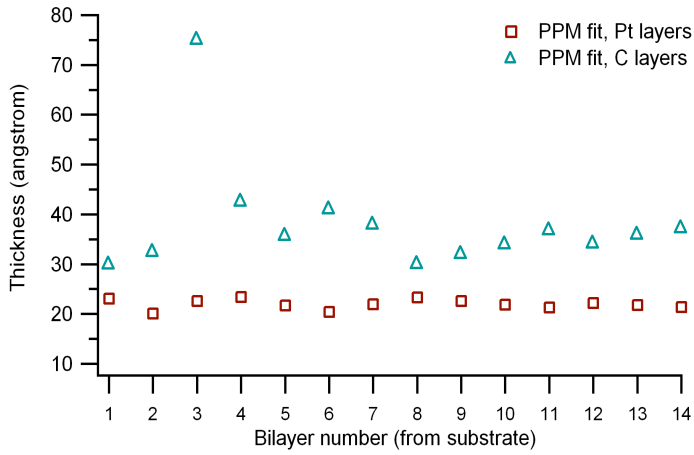 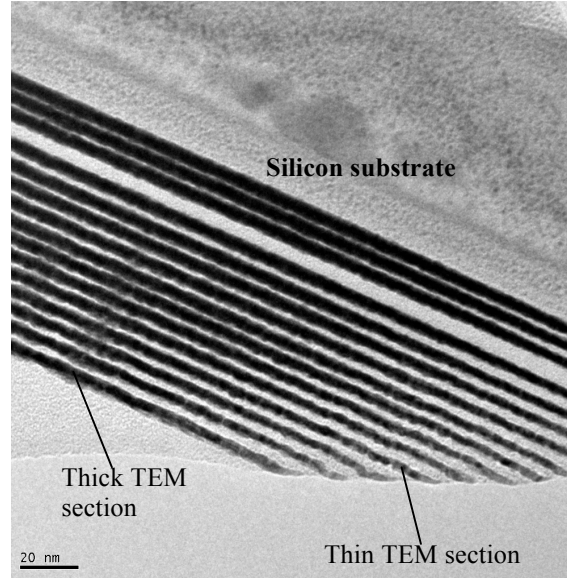

Fig.6: (*left*) the Pt and C thickness trends as derived by PPM. Irregularities in d-spacing are mainly caused by C layers, whereas Pt layers are more regular. The 3rd C layer seems to have an outstandingly large thickness.

(*right*) TEM section of the Pt/C sample: the dark bands are the Pt layers, the white ones are the C layers. The growth direction is the top-down direction in the image. The layers thickness appears to be a function of the section thickness as an effect of the superposition of Pt layer roughness topographies, thus, the PPM results should be compared to the TEM results at the border of the sample. Notice the outstanding thickness of the 3rd C layer, caused by an instability in the e-beam gun used to evaporate the sample.

The verification of the PPM results with a TEM section of the sample confirms at first sight the presence of an outstandingly thick Carbon layer at the expected position (the big bright band in fig. 6 - right). For a more detailed comparison, we extracted and compared individual Pt and C thicknesses obtained from TEM images and PPM analysis. However, thicknesses worked out from TEM profiles can be affected by a significant error, brought in by the superposition of rough Pt profiles on the image plane, causing the Pt layers to appear thicker at the expense of the C layers. The shading effect is larger if the TEM sample is thicker, because a larger amount of profiles are projected on the TEM image. This is easily seen in fig. 6 (right): near the sample edge, where the sample is very thin, the Pt layers are thinner than C layers, whereas this ratio is inverted far from the edge, where the TEM sample is thicker. The Pt layers appear denser and thicker while the C layers appear reduced of the same amount.

As multilayer reflection results from multiple interference events of X-rays at the *mean* interface of each Pt/C and C/Pt couple, the "true" thickness values which can be compared to those derived from XRR analysis should be measured where the TEM sample is very thin, i.e. where the overlapping of roughness profiles is negligible. Unfortunately, the sample edge line is not parallel to the growth direction and a direct extraction of the true layer thickness $d_{Pt}$ from TEM is not possible. We can, indeed, estimate it subtracting *the peak height of the roughness profiles* from the Pt layer thickness $D_{Pt}$, measured far from the sample border, i.e. where the TEM section is thicker but quite uniform. The same value is to be added to the Carbon layer thickness. More precisely:

$$d_{Pt} = D_{Pt} - \sqrt{2}(\sigma_{Pt} + \sigma_C) \tag{1}$$

$$d_C = D_C + \sqrt{2}(\sigma_{Pt} + \sigma_C) \tag{2}$$

The sum of two σ values takes into account the apparent increase/decrease of the layer from both sides, and the factor √2 is the peak /rms ratio. Using the roughness rms values found from the fit (5.3 and 4.2 Å) we obtain that 12.6 Å must be subtracted from the Pt layers measured by TEM and the C layers have to be increased by the same quantity. The comparison between PPM results and TEM *corrected* results is presented in fig. 7 (left), where the TEM results are an average of equally spaced profiles in a region of apparently uniform thickness: the error bars include also the intrinsical TEM uncertainty (5 Å). The average allows to rule out local undulation effects. The agreement is satisfactory.

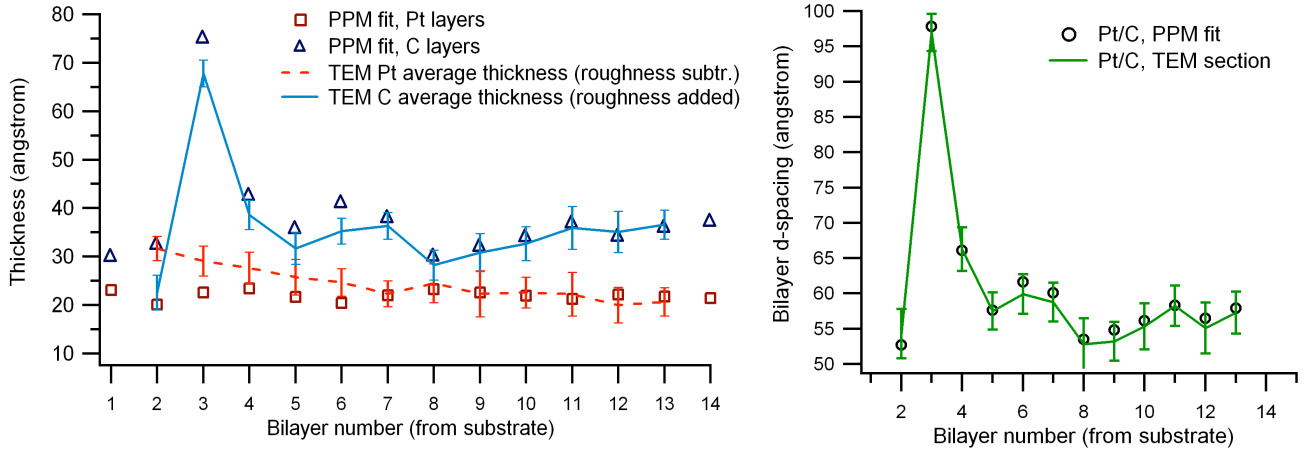

Fig. 7: (*left*) comparison of PPM results (markers) and TEM (lines) after removal of roughness effect from TEM results. The TEM results are obtained from the average of 11 equally spaced profiles of TEM section. Notice the symmetry of discrepancies, which are due to a non-uniform projected roughness in TEM. The error bars of TEM mark the 1 σ confidence limit.

(*right*) *bilayer* d-spacing distribution as calculated by PPM in order to fit the reflectance scan at 8.05 keV. Uncorrected d-spacing values are in agreement with thickness values derived from the TEM analysis within 1.6 Å, i.e. well inside the TEM confidence interval.

It should be noted that the residual discrepancy PPM/TEM (in the first 7 bilayers) could be due to the non-uniform thickness of the sample: for instance, the TEM sample is thicker in the first bilayers, and the correction has probably been underestimated.

A further confirmation comes from the *bilayer d-spacing distribution*, that should not be affected by this correction, since the corrections for Pt and C have opposite sign and so they cancel out when they are summed up to compute the bilayer d-spacing. The comparison of the *bilayer* d-spacing as determined via TEM and through the PPM analysis is shown in fig. 6 (right). The agreement of the two methods is apparent.

### Ni/C multilayer

The Ni/C multilayer (19 bilayers) stack had been initially[19] modeled by a sequence of 6 blocks formed by 3 bilayers; in each block the bilayers were allowed to exhibit a linearly drifting thickness to account for possible short-range thickness fluctuations. The Ni and C densities were assumed to be constant along the stack, whereas the values of the Ni and C roughness were left free to vary in every block. Even if a good fit result was obtained, the XRR best-fit curve showed oscillations not present in the experimental scan: moreover, the generated d-spacing distribution was apparently in disagreement with that resulting from the TEM analysis.

We have repeated the PPM analysis assuming a more realistic structure for the multilayer stack, assuming all the layer thickness values to be free parameters, like for the Pt/C multilayer. Ni and C densities were still kept constant (the assumed Ni density is 8.4 g/cm$^3$ and that of C is 1.6 g/cm$^3$), and the roughness values are still block-structured in order to account for a roughness drift through the stack. The model structure has been optimized to match simultaneously the reflectivity curves at both photon energies, 8.05 and 17.45 keV.

The agreement of the fitting model with the experimental curve is apparent in figs. 8 and 9. The reflectance features are now fitted accurately. The logarithmic $\chi^2$ at 8.05 keV has now a value of 0.61 vs. the much larger value of 2.72 which was obtained with the description based on blocks of bilayers. The $\chi^2$ value at 17.45 keV has changed correspondingly from 1.4 to 0.15.

The main result of this two-energy-scan fit can be summarized as follows: the roughness rms drifts from 3 to 5 Å for both Ni and C, going from the substrate to the multilayer outer surface; the Ni and C layer thickness trend shows (see fig. 10, left) a very limited instability of the Ni layers (a few angstrom) and a very strong variation of the C layers. The main responsible for the overall thickness drift is the imperfect deposition of the C layers. The total d-spacing variation over the entire stack is 2 nm. Rapid variations of C layers could not be detected with the block-based scheme[19], which could return only the average over three bilayers of the true Carbon trend.

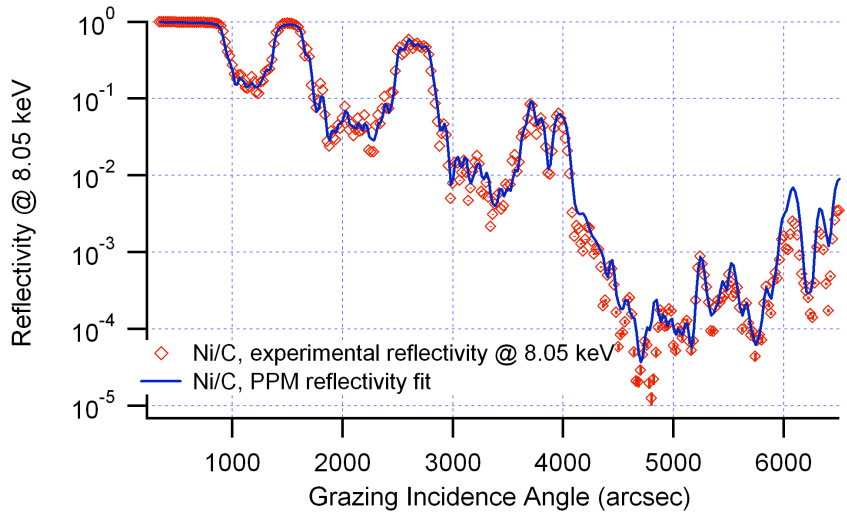

Fig. 8: the reflectivity at 8.05 keV of the Ni/C multilayer sample compared with the fitting model calculated via PPM (log scale). The fit procedure was run simultaneously on both scans at 8.05 and 17.45 keV (s. fig. 9). Error bars (very small) are also included.

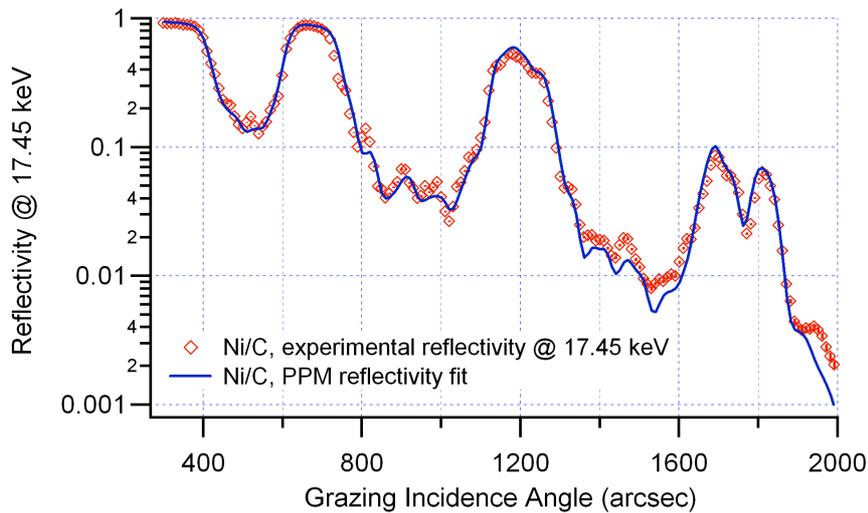

Fig. 9: the reflectivity at 17.45 keV of the Ni/C multilayer sample compared with the fitting model calculated via PPM (log scale). Error bars (very small) are also included.

The layer thickness distribution (fig. 10 - left) has been compared to the data worked out from the TEM section of the same multilayer (fig. 10 – right). We show in fig. 11 (left) the comparison between PPM results (like in fig. 10 -left) and TEM data, corrected for roughness projection: in this case the correction amounts to 12 Å, assuming the maximum roughness value inferred by PPM. The agreement seems to be good only for the first bilayers, for the others the required roughness would have to be much larger than the measured one. This can be explained by observing that the TEM sample is thicker in the image upper part (this can be seen also from the image darkening color in fig. 10 - right). The growth direction goes from the bottom to the top of the image, so both sample thickening and roughness growth contribute to the apparent increase of the Ni layers thickness as the distance from substrate increases. Moreover, imperfections in alignment between the e-beam and multilayer interfaces could also cause a partial shading of the C layers. The agreement between PPM and TEM is, instead, very satisfactory when we compare the bilayer d-spacing values (see fig. 11 - right), which are not affected by the roughness shading effect, as mentioned before (see fig. 7). This result confirms the correctness of the PPM fitting and the interpretation of thickness data from TEM profiles.

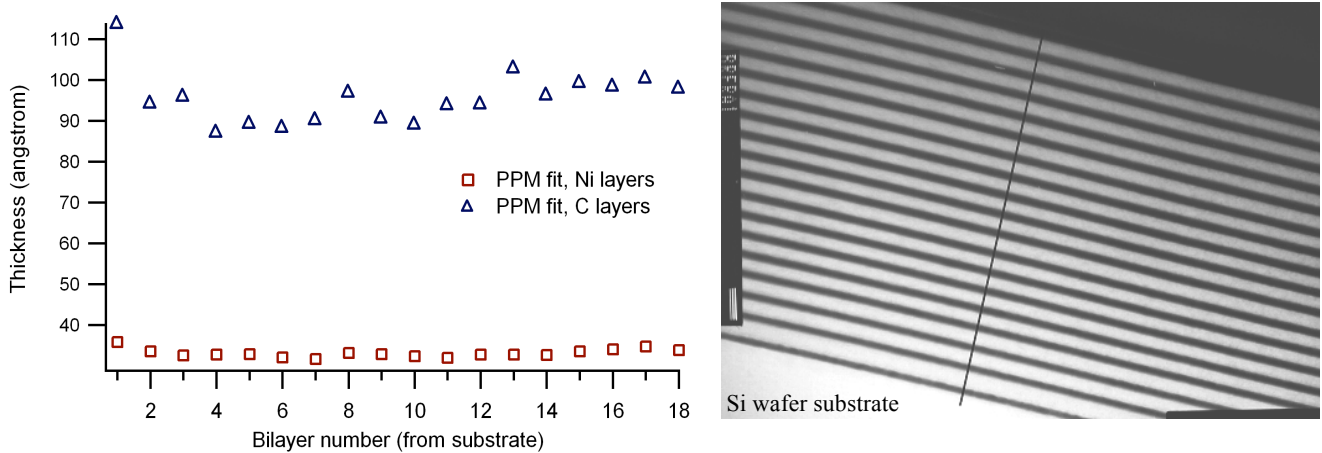

Fig. 10: *(left)* the Ni and C distribution obtained by PPM. Irregularities of the d-spacing are mainly caused by C layers, whereas Ni layers are more regular. The same result was obtained in a previous work, with the stack being schematized as a sequence of blocks: the short-period variation could not be seen in that case.

*(right)* the TEM section of the Ni/ C multilayer. The Ni layers are the dark bands; the C layers are the bright bands: the growth direction is the bottom-up direction. The length of the black marker is 219 nm.

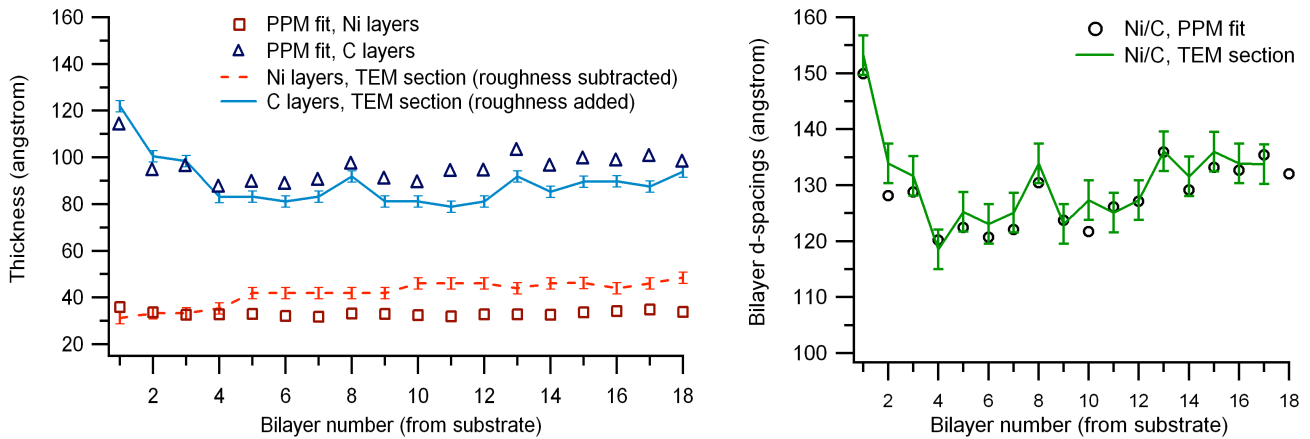

Fig. 11 : *(left)* comparison between TEM and PPM results for the Ni and C thickness distribution. Although the trends obtained with the two methods are similar (notice the matching of peaks in Carbon layers), the discrepancy is larger at the multilayer surface, as a likely result of the roughness/section thickness increase.

*(right)* the bilayer d-spacing distribution in the multilayer as from the PPM optimization. The data matching is satisfactory. The last point of TEM is not considered because the last Nickel layer thickness measurement was not possible.

## 5. DISCUSSION AND APPLICATION TO WIDE-BAND X-RAY MULTILAYERS

The fitting capabilities of PPM have been tested and validated for two quasi-periodic multilayer samples (Pt/C and Ni/C) with significant d-spacing irregularities, obtaining high quality fits at two photon energies simultaneously. Detailed stack models were derived from fitting the corresponding X-ray reflectance curves. The obtained d-spacing distributions are in good agreement with TEM data. The comparison of the latter with the PPM results shows, moreover, that the multilayer Γ ratio can be altered in TEM images by geometrical effects like the superposition of the multilayer roughness profiles in the image plane. As XRR analysis is not affected by image artifacts, the PPM analysis has been able to return the nearly real layers thickness distribution in the multilayer stack.

The fitting capabilities of PPM have been extended to encompass graded multilayers for wide-band reflection X-ray telescopes. Here we present a test performed with a double-graded W/Si multilayer deposited by magnetron sputtering at the *Harvard-Smithsonian Center for Astrophysics* on a fused silica superpolished sample ($\sigma \approx 1$ Å) using a facility developed to deposit multilayers on preformed mirror shells[14]. This sample was deposited during a mirror shell coating run in order to disentangle the intrinsic roughness deposition process from the initial substrate roughness. The multilayer coating structure is formed by two stacks of 20 and 75 bilayers, in which the d-spacing follows the well-known power-law[5] behavior $d(k) = a(k+b)^{-c}$. The 20 outer bilayers are thicker and reflect the softer X-rays in a continuum band. The other 75 bilayers are thinner and are devoted to the reflection of hardest X-rays.

The performed XRR scan at 8.05 keV in fig. 12 has been analyzed with PPM: in this preliminary investigation phase we did not consider each layer thickness as an independent variable, but we allowed the a, b, c parameters in each power law to vary freely, in order to account for a long-term drift of the deposition rate. We allowed also the Γ factor to vary by treating the W and Si layers as distinct power law trends. Roughness rms values are also varying freely and are considered constant throughout the stack. In spite of the small number (12) of free parameters, a very good fit was returned. The fine tuning of the parameters a, b, c for W and Si in the two stacks allows a detailed multilayer structure description and the evaluation of the deposition process reliability and repeatability.

Future developments will be aimed to the application of PPM to XRR scans performed on multilayers deposited using other techniques (e.g. RF magnetron sputtering) and/or with a larger number of bilayers. The comparison of the related results with TEM images will cast light on the reliability of PPM as a diagnostic tool for multilayers for X-ray telescopes.

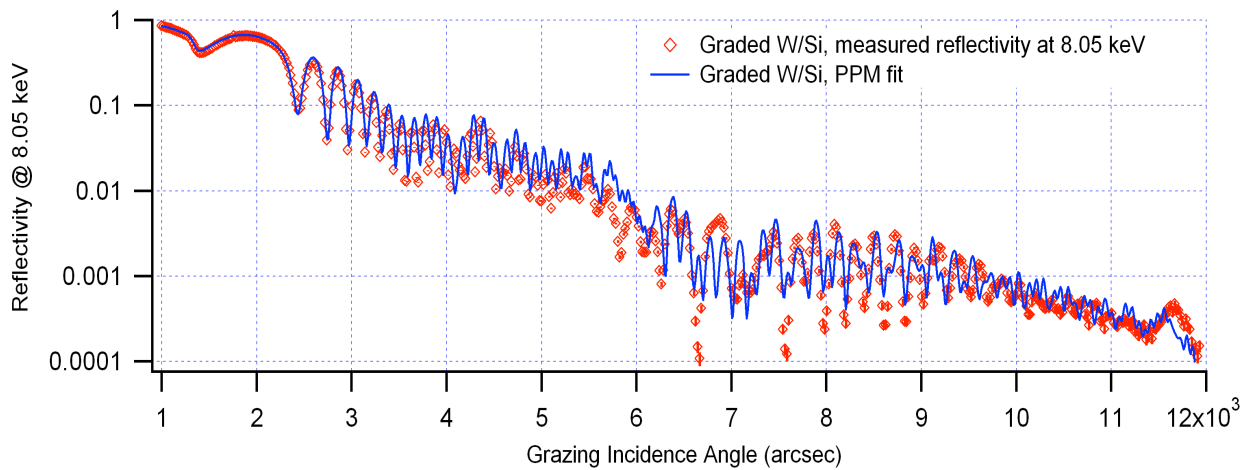

Fig. 12: XRR scan at 8.05 keV of a double graded multilayer W/Si with 20 + 75 bilayers deposited by magnetron sputtering on a fused silica sample at the *Harvard-Smithsonian Center for Astrophysics*.

## ACKNOWLEDGMENTS

The authors gratefully acknowledge the support of O. Citterio, T. Maccacaro (*INAF / Osservatorio Astronomico di Brera*), G. Nocerino *(Media-Lario Techn., Bosisio Parini, Italy)* to this work. They thank S. Romaine, P. Gorenstein, R. Bruni (*Harvard-Smithsonian CfA*, Boston, USA) for the valuable collaboration and in particular for providing them with the W/Si multilayer sample. D. Spiga is indebted with *MURST* (the Italian Ministry for Universities) for the COFIN grant addressed to the development of multilayer coatings for X-ray telescopes, and the European Science Foundation - *COST cooperation, action P7* for the grant which made the PPM training at the ESRF possible. He thanks especially T. Krist (*Hahn-Meitner Institut*, Germany) and M. Idir (*Synchrotron Soleil*, France) for their support. The valuable collaboration of G. Valsecchi, G. Grisoni, M. Cassanelli (*Media-Lario Techn.*) is kindly acknowledged. V. Cotroneo is grateful to *Cariplo Foundation* for the grant funding his fellowship.

PPM is open-source software; it can be freely downloaded from *ftp://www.esrf.fr/pub/scisoft/ESRF_sw/linux_i386_03/*.